\documentclass[aps,pra,tightenlines,twocolumn,showpacs,groupedaddress,superscriptaddress,longbibliography]{revtex4-1}
\usepackage[english]{babel}
\usepackage{float}
\usepackage{graphicx}
\usepackage[caption=false]{subfig}
\usepackage{amsmath}
\usepackage{amsfonts}
\usepackage{amssymb}
\usepackage{mathbrack}
\usepackage{mathtools}
\begin{document}
\raggedbottom

\title{Evolution of Discrete Symmetries}

 \author{Peter Schmelcher}
  \email{Peter.Schmelcher@physnet.uni-hamburg.de}
 \affiliation{Zentrum f\"ur Optische Quantentechnologien, Universit\"at Hamburg, Luruper Chaussee 149, 22761 Hamburg, Germany}
 \affiliation{The Hamburg Centre for Ultrafast Imaging, Universit\"at Hamburg, Luruper Chaussee 149, 22761 Hamburg, Germany}

\date{\today}

\begin{abstract}
Symmetries are known to dictate important physical properties and can be used as a design
principle in particular in wave physics, including wave structures and the resulting propagation dynamics.
Local symmetries, in the sense of a symmetry that holds only in a finite domain of space, can be either
the result of a self-organization process or a structural ingredient into a synthetically prepared physical
system. Applying local symmetry operations to extend a given finite chain we show that the resulting one-dimensional lattice
consists of a transient followed by a subsequent periodic behaviour. Due to the fact that, by construction, the implanted
local symmetries strongly overlap the resulting lattice possesses a dense skeleton of such symmetries.
We proof this behaviour on the basis of a class of local symmetry operations allowing us to conclude upon
the 'asymptotic' properties such as the final period, decomposition of the unit-cell and the length and decomposition of the
transient. As an example case, we explore the corresponding tight-binding Hamiltonians. Their energy eigenvalue
spectra and eigenstates are analyzed in some detail, showing in particular the strong variability of the localization
properties of the eigenstates due to the presence of a plethora of local symmetries.
\end{abstract}

\maketitle

\section{Introduction} \label{sec:introduction}

Symmetries play a prominent role in many branches of physics. They represent a cornerstone both for the analysis
and design of physical systems. Knowing the underlying symmetries of a setup under investigation allows us to predict
certain symmetry-related properties without solving explicitly the corresponding equations of motion. The group theoretical
description and classification of symmetries \cite{Hamermesh62,Cornwell84} , be it discrete or continuous ones,
provides us with e.g. the parity of eigenstates or their rotational quantum numbers and multiplet structure
and resultingly the selection rules for their electromagnetic transitions in atoms, molecules or bulk systems 
\cite{Friedrich17, Bunker79,Girvin19}. The typical and widely accepted situation assumes that a certain spatial
symmetry holds for all of the space covered by the physical system under investigation, in short, it represent a global symmetry.
The rotational symmetry of atoms, the point group symmetries of molecules and the translation group
symmetries of crystals all belong to this case. This is, however, by far not the most general structural behaviour
and many more complex but still geometrically appealing physical systems do not possess any global symmetry. One prominent
example are quasicrystals \cite{Levine84,Shechtman84,Suck02} who generically do not possess any global symmetry
but are governed by a plethora of local symmetries arranged in a quasiperiodic manner \cite{Morfonios14,Roentgen19}.
Here the notion of a local symmetry refers to the situation where a symmetry holds only on a limited subdomain
of the domain of definition of a system. Quasicrystals represent one class of systems on the rich transition
route from periodicity to disorder and are nowadays of major importance in material science and technology \cite{Macia06,Macia12}.

More recently the concept of local symmetries has been further developed and applied to reveal a number of
physical properties and phenomena. Breaking e.g. the symmetry of a crystalline translation group and remaining
only with a local symmetry has been shown to lead to constants of motion which represent non-local currents
that generalize the Bloch theorem for periodic crystals \cite{Kalozoumis14a}. Multiple local symmetries can be combined
to provide setups that are characterized by so-called non-gapped or gapped local symmetries, where the gap 
refers to the space between the appearance of local symmetries, or systems with complete local symmetries
for which local symmetries cover the setup completely. Applications to wave systems of acoustic \cite{Kalozoumis15}
or electromagnetic \cite{Schmitt20} origin have shown that perfectly transmitting resonances can be
designed based on sum rules of the non-local currents \cite{Kalozoumis14b} which can also be used
to detect local order via the analysis of the wave field. Indeed, the non-local currents fulfill generalized continuity
equations in the framework of both the discrete \cite{Morfonios17} and the continuous theory \cite{Schmelcher17}.
A corresponding computational approach to efficiently handle locally symmetric wave systems has been
developed in ref.\cite{Zambetakis16}. Remarkably in ref.\cite{Morfonios20} it has been shown that the
presence of local reflection symmetries in a one-dimensional finite and disordered chain severely impacts
wave localization and transport. Due to the local symmetries correlations are imprinted into the wave
field and specifically the corresponding transfer can be significantly enhanced if the ocurring local 
symmetries overlap with respect to their domains.

In the present work we follow a different pathway and explore a new class of chains that are neither globally
symmetric nor disordered and also not locally symmetric in the above sense of a concatenation of subchains with
global symmetries. Our strategy is to generate a chain based on an initial seed that consists of a finite number of elements 
by the consecutive application of local symmetry operations, namely reflections, according to a given rule.
This way a chain is designed that exhibits and is completely covered by a sequence of {\it{overlapping}} domains with
local symmetries. We demonstrate the plethora of possible evolutionary sequences achievable by the repeated
application of local symmetry operations. In general, and starting from the seed sequence, the chain evolves
(in space) until it finally, following a finite transient, becomes periodic. We provide a general proof of
this behaviour which is of constructive character and allows us to predict the transient to periodicity depending
on the applied local symmetry transformation. It also yields all relevant quantities such as the final period
and the length of the transient. A plethora of transients and resulting periodic behaviour can be established
this way. In a second step we translate the obtained scheme and chains at hand of 
specific examples to a tight-binding (TB) Hamiltonian which we subsequently diagonalize. The resulting energy
eigenspectrum and eigenstates are then analyzed in some detail. We observe a rich spectral behaviour which
changes signficantly for varying local symmetry transformations. Strong localization and delocalization behaviour 
determined by the presence of local symmetries are detected in the eigenstate properties and are interspersed
into series of eigenstates with a smooth energy dependence of their spread. 

We proceed as follows. In section \ref{sec:lsdchains} we first introduce our concept of local symmetry operations
and dynamics to generate chains with many overlapping domains of local symmetries. We do so by using some representative
examples. This allows us already to showcase the existence of a transient and the resulting periodic behaviour which
vary substantially with varying transformations. We then provide a general constructive proof of the overall behaviour
of the local symmetry dynamics generated chains by full induction. In section \ref{sec:tbh} we map the previously
obtained chains onto a TB Hamiltonian and discuss the resulting energy spectra as well as the so-called
eigenstate maps. In section \ref{sec:concl} we provide our conclusions and an outlook.

\begin{table*}[t]
  \begin{center}
    \begin{tabular}{|l|l|} 
	    \hline
      \textbf{Seed} & ABCDEFG \\
      \hline
	    7:1 & $|_7$ G \underline{FEDCBA $|_1$ A $|_7$ AABCDEF $|_1$ F $|_7$ F} FEDCBA ....  \\
      \hline
	    7:2 & $|_7$ GF \underline{EDCBA $|_2$ AB $|_7$ BAABCDE $|_2$ ED $|_7$ DE} EDCBA $|_2$ AB $|_7$ BAABCDE....\\
      \hline
	    7:3 & $|_7$ GFE \underline{DCBA $|_3$ ABC $|_7$ CBAABCD $|_3$ DCB $|_7$ BCD} DCBA $|_3$ ABC $|_7$ CBAABCD....\\
      \hline
	    7:4 & $|_7$ GFEDCBA $|_4$ ABCD $|_7$ D \underline{CBAABC $|_4$ CBAA $|_7$ AABCCBA $|_4$ ABCC $|_7$ C} CBAABC....\\
      \hline
	    7:5 & $|_7$ GFEDCBA $|_5$ ABCDE $|_7$ EDCBAAB $|_5$ BAABC $|_7$ C \underline{BAABBA $|_5$ ABBAA $|_7$ AABBAAB}\\
		& \underline{$|_5$ BAABB $|_7$ B} BAABBA....\\
      \hline
	    7:6 & $|_7$ GFEDCBA $|_6$ ABCDEF $|_7$ FEDCBAA $|_6$ AABCDE $|_7$ EDCBAAA $|_6$ AAABCD $|_7$ DCBAAAA\\
		& $|_6$ AAAABC $|_7$ CBAAAAA $|_6$ AAAAAB $|_7$ B \underline{A}AAAAA....\\
      \hline
    \end{tabular} 
    \caption{The seed of the chain and the local symmetry generated transients as well as periodic final behaviour for 
          specific examples based on the local symmetry dynamics rules 7:1 to 7:6. The abbreviation $|_k$ stands for a reflection operation
	  at the indicated position which reflects $k$ elements of the symbolic code to the left of this position. The
	  underlined sequences represent for each LSD rule the unit-cell of the final periodic behaviour of the chain.}
    \label{tab:table1}
  \end{center}
\end{table*}

\section{Local symmetry dynamics generated chains} \label{sec:lsdchains}

This section is dedicated to the introduction of our concept of local symmetry dynamics based on the
consecutive application of local symmetry operations in one spatial dimension. After explaining the basics we provide a specific
example in order to gain some intuition with respect to the evolution of the underlying symbolic code
including the transient and the final periodic behaviour. As a next step we will provide a constructive
proof for the general case which allows us to extract the relevant properties and behaviour for an
arbitrary case.

\subsection{Symbolic code and local symmetry operations: Basic concept and examples} \label{subsec:sclso}

Discrete reflection and translation symmetries are ubiquitous in atomic, molecular and crystalline systems, respectively.
Quasicrystals with their aperiodic order host these symmetries only in a local sense i.e. only in certain subdomains of the 
quasiperiodic chain a, for example, reflection symmetry can be found. While there is many such domains in an aperiodic
systems with long-range order \cite{Morfonios14}, they are typically disconnected in the sense that they are not generated
one from the other. Our starting-point is therefore the idea of creating chains with a large number of local symmetries on overlapping
domains based on the repeated application of local symmetry operations. These chains will neither be globally symmetric, i.e.
neither reflection symmetric nor periodic, but will exhibit (see below) an evolution of their local symmetries in the course
of the multiple application of the local symmetry operations. We call this evolution and approach in the following local
symmetry dynamics (LSD). Since this is a general conceptual approach independent of a specific physical platform we will 
develop it on the level of a symbolic code. Later on (see section \ref{sec:tbh}) a concrete realization in the form
of a TB Hamiltonian will be investigated.

The key ingredients read as follows. We start from a seed sequence which represents the initial segment of our final symbolic 
code. Based on this seed we apply at the end of it a local reflection operation of $n$ symbols or elements occuring to the left of
the position of the corresponding reflection 'axis'.  Subsequently, we apply another local reflection operation of $m$ symbols
at the end of the previously obtained chain. We then repeat this procedure again and again which is encapsulated in the
$n:m$ LSD rule. By construction this guarantees that the local symmetries present in the generated chain are strongly
overlapping. To get an impression of how such a chain develops and in order to gain some intuition about the evolution
of the corresponding local symmetries we will inspect the case of a series of specific examples namely $n=7,m=1,...,6$ in the following.

Table \ref{tab:table1} shows the evolution of the symbolic code for the above-mentioned cases based on the seed sequence
ABCDEFG. $|_k$ indicates the position of the local reflection axis meaning that $k$ chain elements to the left of it 
are reflected to the right. We observe that in all cases the LSD leads from an initial seed via a transient to a final periodic
periodic behaviour of the chain. Pictorially speaking the seed undergoes a metamorphosis thereby gradually increasing
the degree of the symmetry contained in the chain until it finally becomes periodic. Obviously, the number of different
symbols is decreasing along the chain in the course of the evolution to periodicity (see subsection \ref{subsec:proof}
for quantitative statements along these lines). In this vein the LSD leads to a well-defined loss of information 
from the seed to the asymptotic behaviour of the chain.

Let us discuss the different cases in some more detail. For the cases 7:1, 7:2, 7:3 the transient consists of only
$8,9$ respectively $10$ symbols, i.e. it is very short. The final unit-cell of the periodic sequence is comparatively long amounting
to $16,18$ and $20$ elements. These unit-cells exhibit a plethora of 'internal' local symmetries being
local reflections and translations. This situation changes when moving on to the cases 7:4,7:5 and 7:6. Here the
transients become substantially longer amounting to $19, 32$ and $73$ elements, respectively. The final unit-cells contain
$22,24$ and a single elements. While for the case 7:4 only the A,B and C elements are asymptotically present it is only
A which occurs finally for the case 7:6. With increasing value of $m$ the number of local symmetries increases for the unit-cell
of the asymptotically emerging periodic behaviour. Having gained an intuition of the 'phenomenology' that occurs in
case of our specific example, the question arises how representative this behaviour is, and whether one can 
derive and understand the above-mentioned properties in the general case $n:m$ quantitatively.
To this end, we will provide a proof by explicite construction and full induction in the following subsection.

        \begin{figure*}[t]
            \centering
\hspace*{-1.4cm}
\includegraphics[width=1.4\textwidth]{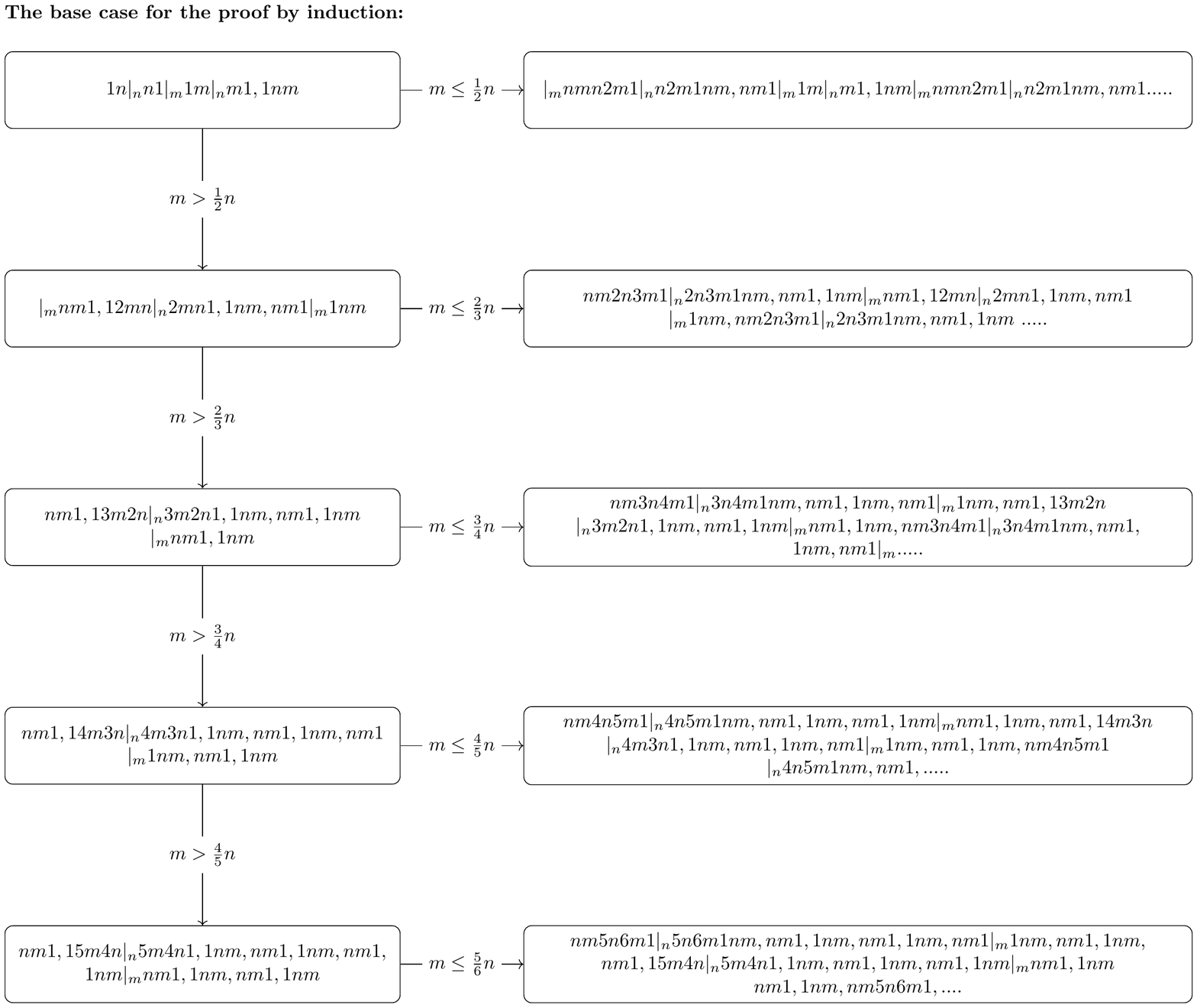} \\ \vspace*{-1.5cm}
\caption{The initial steps for the proof by induction. The initial seed is contained in the uppermost left rounded box.
The flowchart shows (i) the pathway to periodic behaviour along the horizontal branches and (ii) the pathway to higher
values of $m$ passing along the vertical branching sequences. A branching occurs for each case $m = \frac{p}{p+1} n, p = 1,2,...$.
Abbreviations are: $1n = \{1,...,n \}$, $n1 = \{n,....,1\}$, $1nm = \{1,...,n-m\}$, $nm1 = \{n-m,...,1\}$,
$1m = \{1,...,m\}$, $m1 = \{m,...,1\}$, $12mn = \{1,...,(2m-n)\}$, $2mn1 = \{(2m-n,...,1\}$ and
more generally $nmpn(p+1)m1 = \{(n-m),...,(pn-(p+1)m+1)\}$. $|_n$ indicates a local reflection operation of $n$ symbols.}
\label{fig:proof1} 
        \end{figure*}

        \begin{figure*}[t]
            \centering
\hspace*{-1.4cm}
\includegraphics[width=1.4 \textwidth]{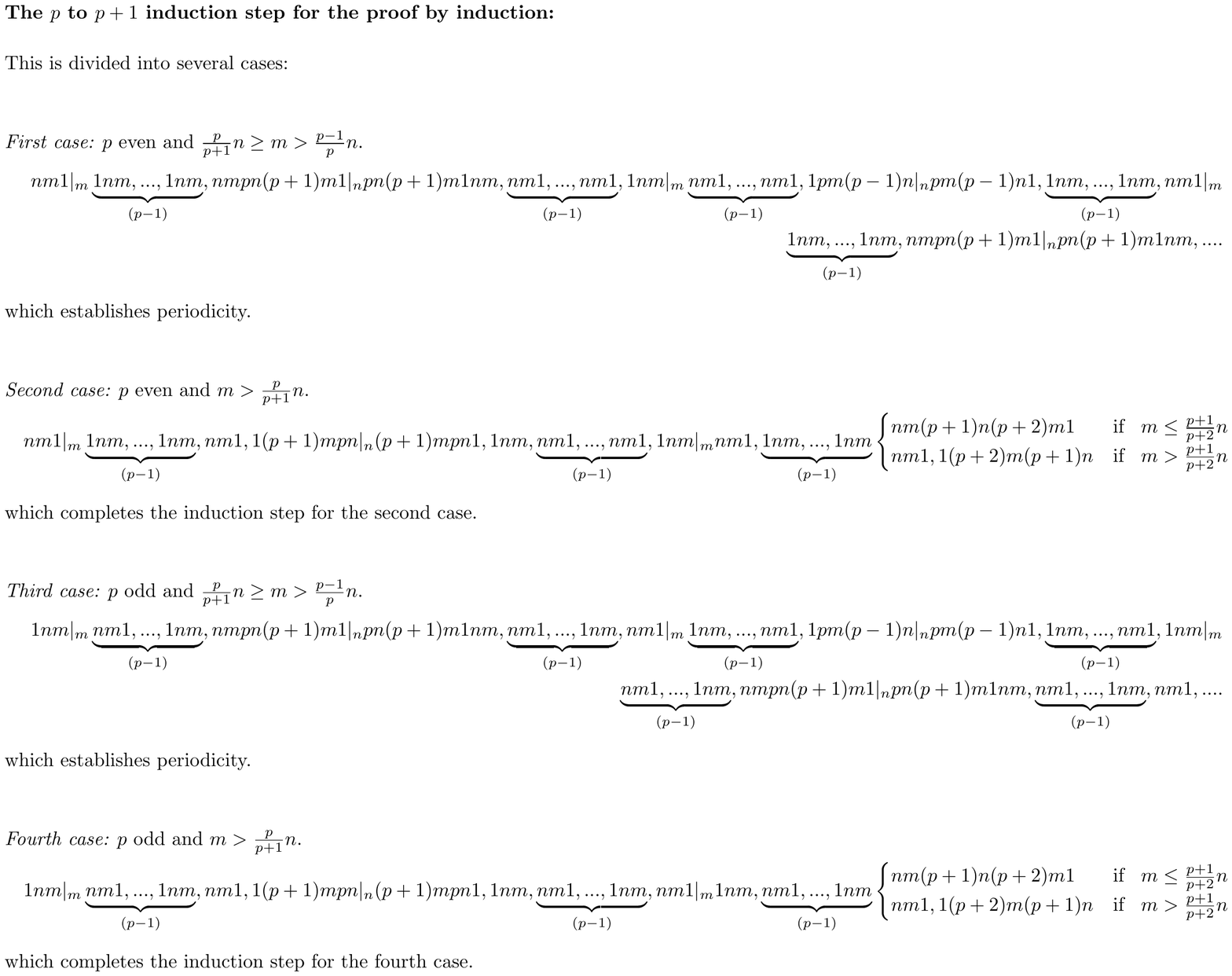} \\ \vspace*{-2cm}
\caption{The $p$ to $p+1$ induction step for the proof involving all possible branchings i.e. cases.
Notation is the same like in Fig. \ref{fig:proof1}.}
\label{fig:proof2}
        \end{figure*}

\subsection{Proof of the general case and resulting properties} \label{subsec:proof}

We provide now a constructive proof by induction of the symbolic sequence generated by a general $n:m$ LSD
where we always assume that $n>m$.
To this end it is advisable and instructive to employ as an encoding of the elements of the chain not the
alphabatical letters used in the previous subsection \ref{subsec:sclso} but a numerical encoding. Specifically this
means that the initial seed ABCD.... is now replaced by the symbolic sequence $1,...,n$ if $n$ seed elements are present
(please note here the double use of $n$; first as a symbolic variable and second as a number).
This way of encoding facilitates the local symmetry operations in the course of the LSD substantially and 
adds to the transparency of the below-given proof in Figs.\ref{fig:proof1} and \ref{fig:proof2}, respectively.
The numbers used here are therefore unique placeholders for a corresponding symbol.
The hereby used notation is as follows: $1n = \{1,...,n \}$, $1nm = \{1,...,n-m\}$, 
$1m = \{1,...,m\}$, $12mn = \{1,...,(2m-n)\}$, and more generally $nmpn(p+1)m1 = \{(n-m),...,(pn-(p+1)m+1)\}$.
$|_n$, as mentioned already above, indicates a local reflection operation on $n$ symbols.

To start the proof by full induction one has to perform some first steps. We will here for reasons of
illustration of the concept and overall flow of the general case provide a few more initial steps as absolutely
necessary for the proof. Fig. \ref{fig:proof1} shows a flowchart of the first few steps and branching sequences.
The flowchart consists of horizontal and vertical flows which emerge at every branching point $m = \frac{1}{2}n,
\frac{2}{3}n, \frac{3}{4}n,\frac{4}{5}n$. Following the corresponding horizontal branches for $m \le \frac{1}{2}n,
\frac{2}{3}n, \frac{3}{4}n,\frac{4}{5}n$ leads to periodicity with different characteristic unit-cells which
can be read off from the r.h.s. boxes in Fig.\ref{fig:proof1}. In case $m$ larger than a given ratio one 
proceeds along the vertical direction of the flowchart until, finally, the horizontal case has to be
chosen, which leads to periodicity. Already from these initial steps it is evident that both the transient
as well as the final unit-cell are composed of $1nm$, 
$nmpn(p+1)m1 = \{((n-m),...,(pn-(p+1)m+1))| p \in \mathbb{N}\}$ and
$1(p+1)mpn = \{(1,...,((p+1)m-pn)\}$ as well as
their inverted sequences. The increasing length of the transient as well as the size of the final unit-cell
with increasing ratio $\frac{m}{n}, m,n \in \mathbb{N}$ is evident from these initial steps already. 


In Fig. \ref{fig:proof2} the $p$ to $p+1$ induction step is performed. The various cases are explicitly
mapped out and the route to periodicity (horizontal branches in Fig.\ref{fig:proof1}) as well as the
extension of the transient (vertical branch in Fig.\ref{fig:proof2}) up to the next branching point are
shown. This concludes the proof by explicit construction and provides us with the content of the transient
and the final unit-cell. In this $p-$th step sequences of $1nm$ (respectively $nm1$) appear in a $p-$fold manner
intermingled with elements of the structure $nmpn(p+1)m1,1(p+1)mpn$. As a result we can now answer
the following questions. What is the length of the symbolic code, i.e. the number of symbols, until
periodicity sets in ? Based on the above proof we can now sum up the lengths of the sequences
in the transient which finally amounts to $2(n+mp+m)$ where $p$ is the last branching point index
(note that the onset of periodicity is here counted from the corresponding branching points on).
If $m$ is of the order of $n$ and consequently $p$ is of the order of $m$ this states that
the length of the transient is of the order $n^2$.
Another question concerns the length of the emerging periodicity i.e. the size of the unit-cell. This 
turns out to simply be $2(n+m)$ in agreement with the above observations. Finally we note that
the unit-cell of the periodic behaviour consists of $n-m$ different symbols out of originally $n$ symbols
in the seed. It should be noted that within a series of LSD rules $n:1,....,n:(n-1)$ not every consecutive branch in the
tree constituted above in the framework of our proof will be realized.

\section{Tight-Binding Hamiltonian: Spectral analysis and Eigenstate Maps} \label{sec:tbh}

To explore the properties of the LSD generated chains derived in the previous section \ref{sec:lsdchains}
we use in the following a mapping onto a corresponding TB Hamiltonian.
For a discrete chain of length $N$ with sites $\{i|i=1,...,N\}$ we assume that there is only a single value for the off-diagonal
couplings $C$ of the nearest neighbors $\langle i,j \rangle$ and the on-site energies $D_i$ follow the LSD chains. This means
each symbol of the symbolic code corresponds to a unique value $D_i$ (see below for more details) of this on-site energy 
on site $i$. We therefore assume that the Hamiltonian takes on the following appearance

\begin{equation}
{\cal{H}} = \sum_{i=1}^{N} D_i |i \rangle \langle i| + \sum_{ \langle i,j \rangle} C |i \rangle \langle j| 
\label{eq1}
\end{equation}

In the following subsections we will analyze in some detail the eigenenergy spectrum and the properties of the
eigenstates for different LSD chains. Before we enter into a discussion of the corresponding results some remarks
are in order concerning the properties of other well-known examples of TB systems. This will allow us
to contextualize the results on our LSD chains.

The simplest case of a TB chain refers to the monomer chain AAA... For both open and periodic boundary
conditions a single band occurs for the spectrum of the energy eigenvalues. This case can be solved analytically
and for periodic boundary conditions (BC) two-fold degeneracies are encountered whereas open BCs
lead to non-degenerate energy levels. For both cases the eigenstates are delocalized (bulk) states.
A chain of dimers ABABAB... results in two bands separated by an energy gap and the periodic case can be solved analytically 
(see refs.\cite{Goringe97,Kouachi06,Kulkarni99,Willms08} for a corresponding analysis allover).
More complicated unit-cells and open boundary conditions lead to multiple bands and in particular to the
presence of gap energy eigenstates which are localized at the edges of the finite chain and whose number
is of the order of the size (number of elements) of the unit-cell, depending on how the unit-cell
is cut off \cite{Zak85,Fukui20}. Individual impurities
add to this spectrum eigenstates which are exponentially localized \cite{Thouless74,Brandes03}.
The situation becomes richer in terms of symmetries in the case of quasicrystals with their long-range aperiodic order 
\cite{Levine84,Shechtman84,Suck02}. Indeed, the iterative action of a given substitution rule underlying aperiodic
lattices lead to a plethora of e.g. local reflection symmetries or, more precisely, 
they lead to a quasiperiodic recurrence of reflection symmetries \cite{Morfonios14} which manifests itself in the
corresponding return map. Depending on the spatial complexity of the aperiodic system under consideration,
the energy eigenvalues cluster into so-called quasibands and the corresponding
eigenstates are neither delocalized allover the bulk nor exponentially localized, but dubbed 'critical'.
Of particular relevance to the following investigation of TB Hamiltonians based on LSD chains
is the local symmetry theory of resonator structures in binary aperiodic chains developed in ref.\cite{Roentgen19}.
For weak inter-site coupling it has been shown that the eigenstate profile is largely determined by
so-called local resonator modes i.e. eigenmodes confined to locally symmetric domains of the chain.
Eigenstates within a given quasiband share the same local resonator modes.
Corresponding edge states are then typically localized on locally symmetric domains at the edges of the
finite chain.

In the following subsections we will analyze the energy eigenvalue spectrum and the corresponding eigenstates
via an eigenstate map for different LSD generated chains, in analogy to the cases considered in section
\ref{subsec:sclso}. The seed $\{A,B,C,D,E,F,G\}$ in table \ref{tab:table1} and the values of the
coefficients $D_i, i=1,...,7$ in eq.(\ref{eq1}) correspond to $\{1,2,3,4,5,6,7\}$ respectively 
whereas the off-diagonal coupling is equal to one. This imprints an intrinsic energy scale into the
TB system with a significant, but not too large, contrast $\frac{(D_{i+1}-D_i)}{C}$. Later on (see
subsection \ref{subsec:sc}) we will consider also chains with correspondingly smaller contrast i.e. for stronger
off-diagonal couplings. For illustrative reasons our chains consist of the transient including the seed as well
as three unit-cells of the emerging periodic behaviour.

\subsection{The 7:1 LSD chain}

\begin{figure}[t]
\hspace*{-1.0cm}\includegraphics[width= 0.8 \columnwidth]{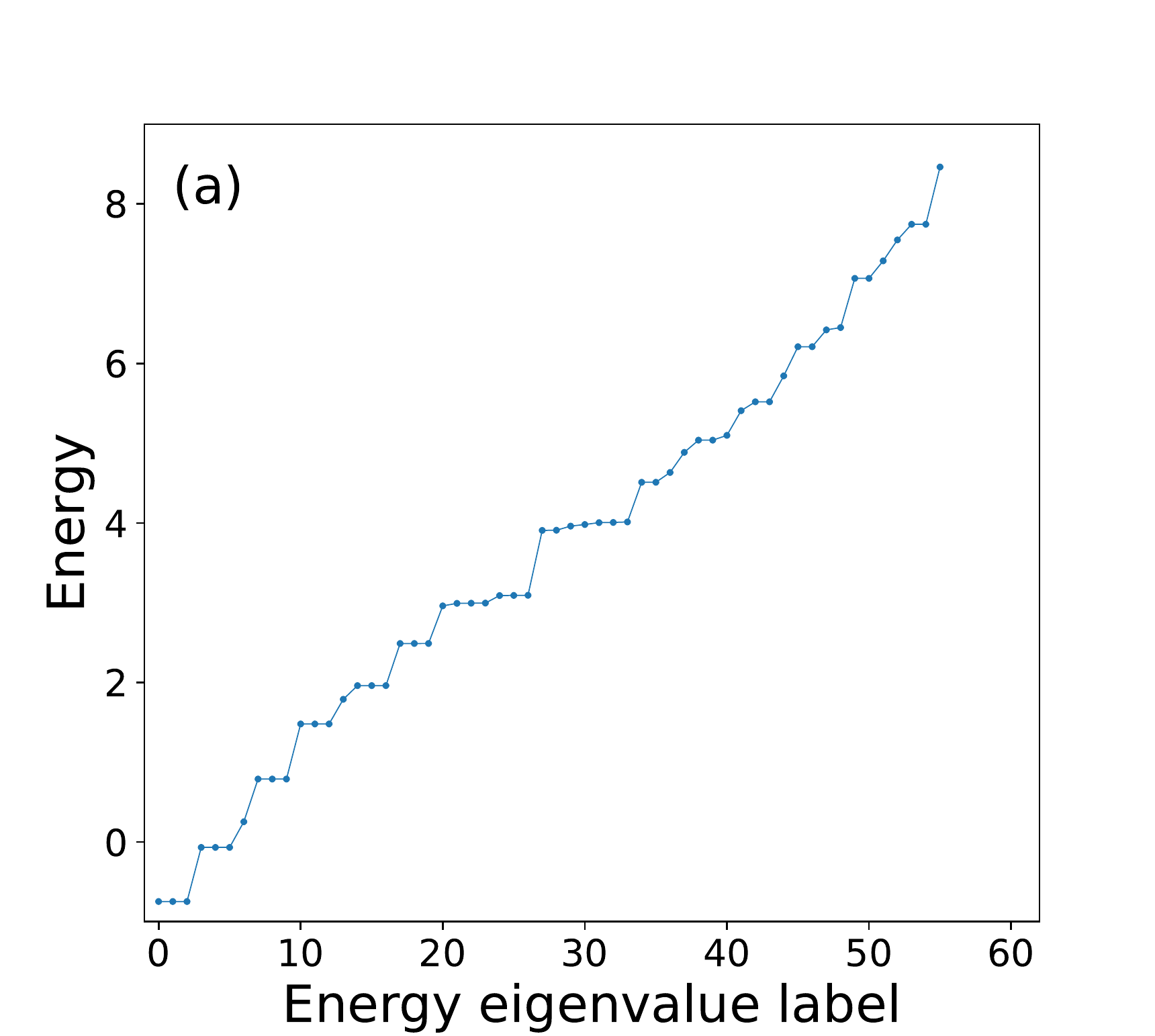} 
\includegraphics[width= \columnwidth]{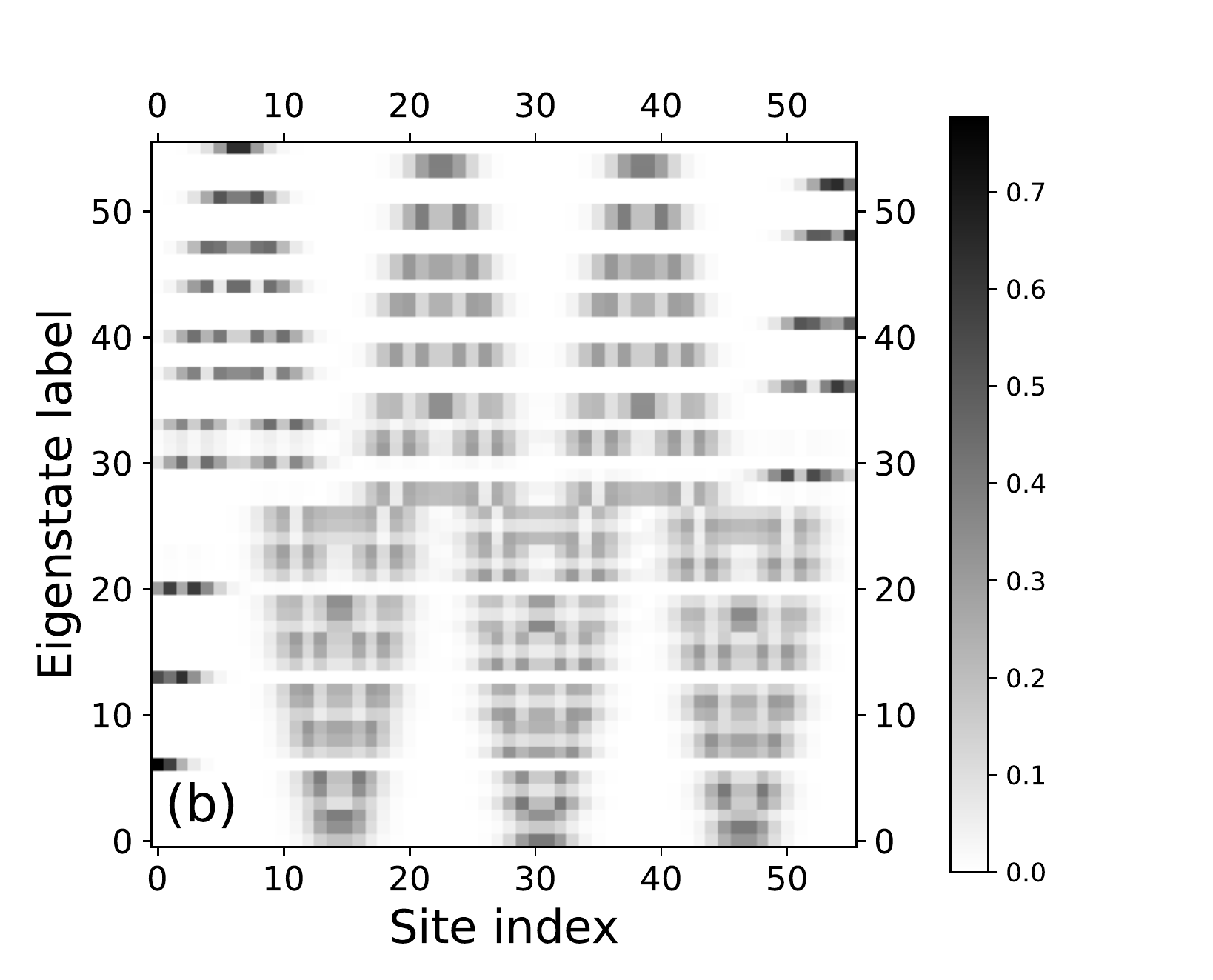} 
\caption{(a) Spectrum of the energy eigenvalues for a TB chain emerging from a seed ABCDEFG and subsequent application
of the 7:1 LSD up to a total of 56 sites, which includes three unit-cells of the final periodic behaviour.
The line is drawn to guide the eye.
(b) The corresponding grey scale eigenstate map showing the magnitude of the amplitudes on all
sites of the chain for all eigenstates whose labeling $0,...,60$ is according to an increasing eigenenergy.
We use open BCs and the on-site energies corresponding to $\{$A,B,C,D,E,F,G$\}$ 
are $\{1,2,3,4,5,6,7\}$ respectively whereas the off-diagonal coupling is equal to one.}
\label{fig:Fig3ab}
\end{figure}

For this case the transient following the seed consists only of eight chain element and is followed
by a periodic behaviour of period 16 with the unit-cell FEDCBAAAABCDEFFF (see table \ref{tab:table1}).
Fig.\ref{fig:Fig3ab}(a) shows the spectrum of the energy eigenvalues for the 7:1 LSD chain.
Clusters of eigenvalues show a plateau-like behaviour separated by gaps.
There is a few states localized in these energy gaps. This behaviour is reminescent of what could be expected
from a finite periodic chain with open boundary conditions. 

Let us now analyze the underlying eigenstates and their localization behaviour in some detail.
The grey scale eigenstate map shown in Fig.\ref{fig:Fig3ab}(b) provides a complete overview of the magnitude
of the amplitudes on all sites of the chain for all eigenstates obtained via diagonalization of the
Hamiltonian (\ref{eq1}). The energetically lowest states 0-2 are predominantly localized
on the three locally symmetric AAAA sequences centered around the sites $14/15, 30/31, 46/47$
in the chain which correspond to the lowest
on-site energies. With increasing degree of excitation these eigenstates become increasingly delocalized but
still centered around the corresponding AAAA sequences (see e.g. states 3-6,8-13, etc.)

Interdispersed into this rather homogeneous sequence of spatially broadening eigenstates
are left-localized (see states $7,13,20$) edge states.
In the middle of the spectrum the AAAA centered states start to overlap. For higher
energies this process is inverted and spatially shifted. Now, the eigenstates are centered
around the high energy locally symmetric FFFF sequences (sites $22/23, 38/39,54/55$) of the chain 
and while they are originally rather 
delocalized they become with increasing degree of excitation more and more focused on the FFFF sequence only. It should 
be noted that some of these FFFF-centered eigenstates are localized on two and others on a single
of these high energy FFFF subdomains. This feature depends on the cutoff of the sequence, i.e.
for another cutoff a localization of some of the states on all three and others on only two can
be observed. Interdispersed into this inverted sequence
are states $30,33,37,40,44,47,51,55$ living on the original seed ABCDEFG and its counterpart which is the first
locally reflected sequence GFEDCBA. With increasing degree of excitation this sequence of states
becomes increasingly focused around the high energy center GG. 

Let us conclude already at this early point of the discussion of our TB analysis
with some observations that will equally hold for the following TB setups.
The LSD generated chain exhibits a diversity of different localization properties of their eigenstates 
which are triggered by two main ingredients. First and foremost the localization is structurally organized by the presence of 
local symmetries and due to the hierarchy of such symmetries in the chain generated by the LSD rules
there is a hierarchy of localization behaviour encountered in the eigenstate profiles with increasing
energies (see ref.\cite{Roentgen19} for a justification and analysis of this property based on
degenerate perturbation theory and a corresponding application to quasiperiodic binary chains).
This behaviour is supported and/or directed by the particular choice of our on-site energies
which provides an energetical order for the occupation of the locally symmetric sequences due to 
the increasing values within the sequence A,...,G.

\subsection{The 7:3 and 7:5 LSD chain}

\begin{figure}[t]
\hspace*{-1.1cm} \includegraphics[width=0.8 \columnwidth]{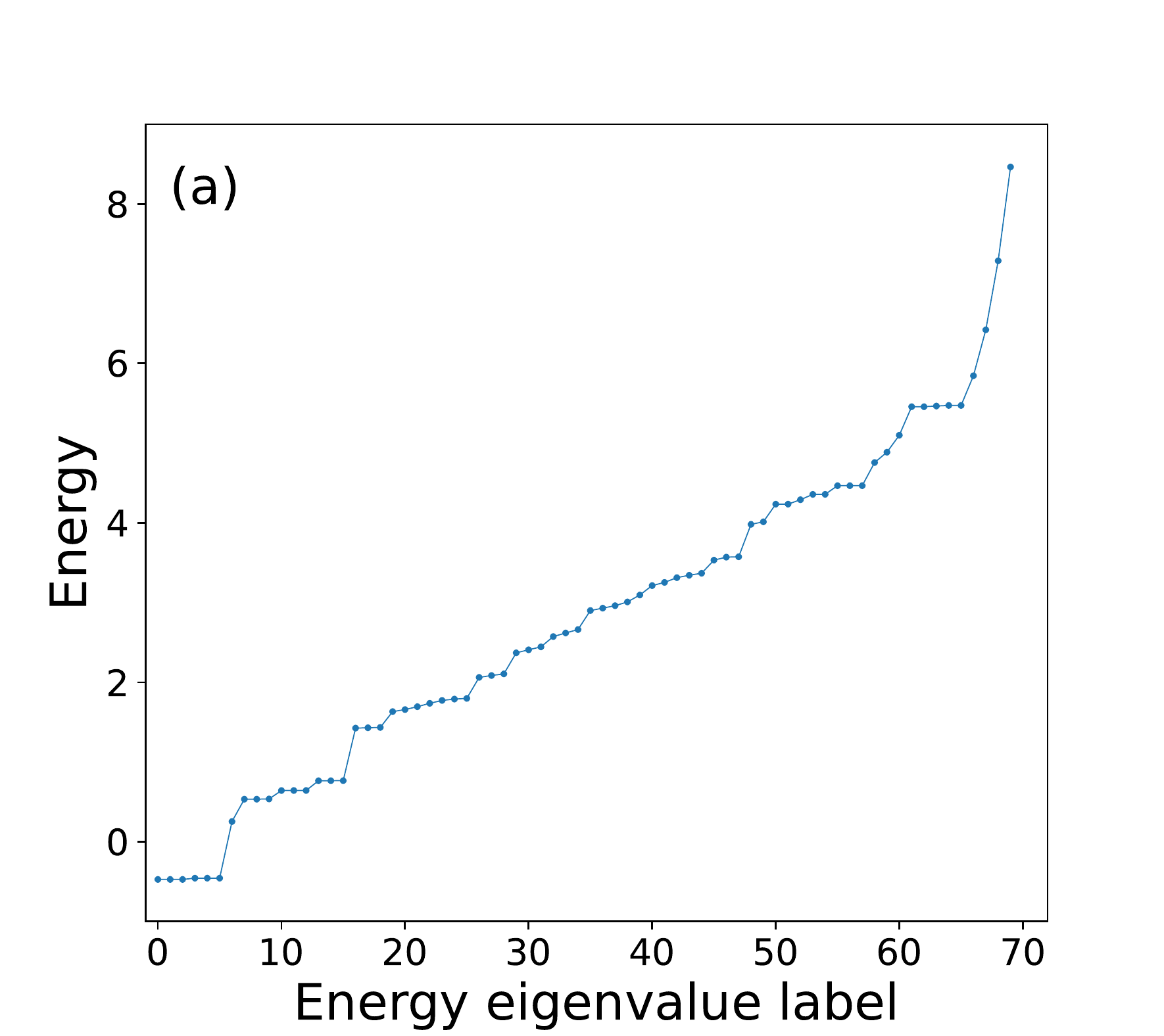} 
\includegraphics[width=\columnwidth]{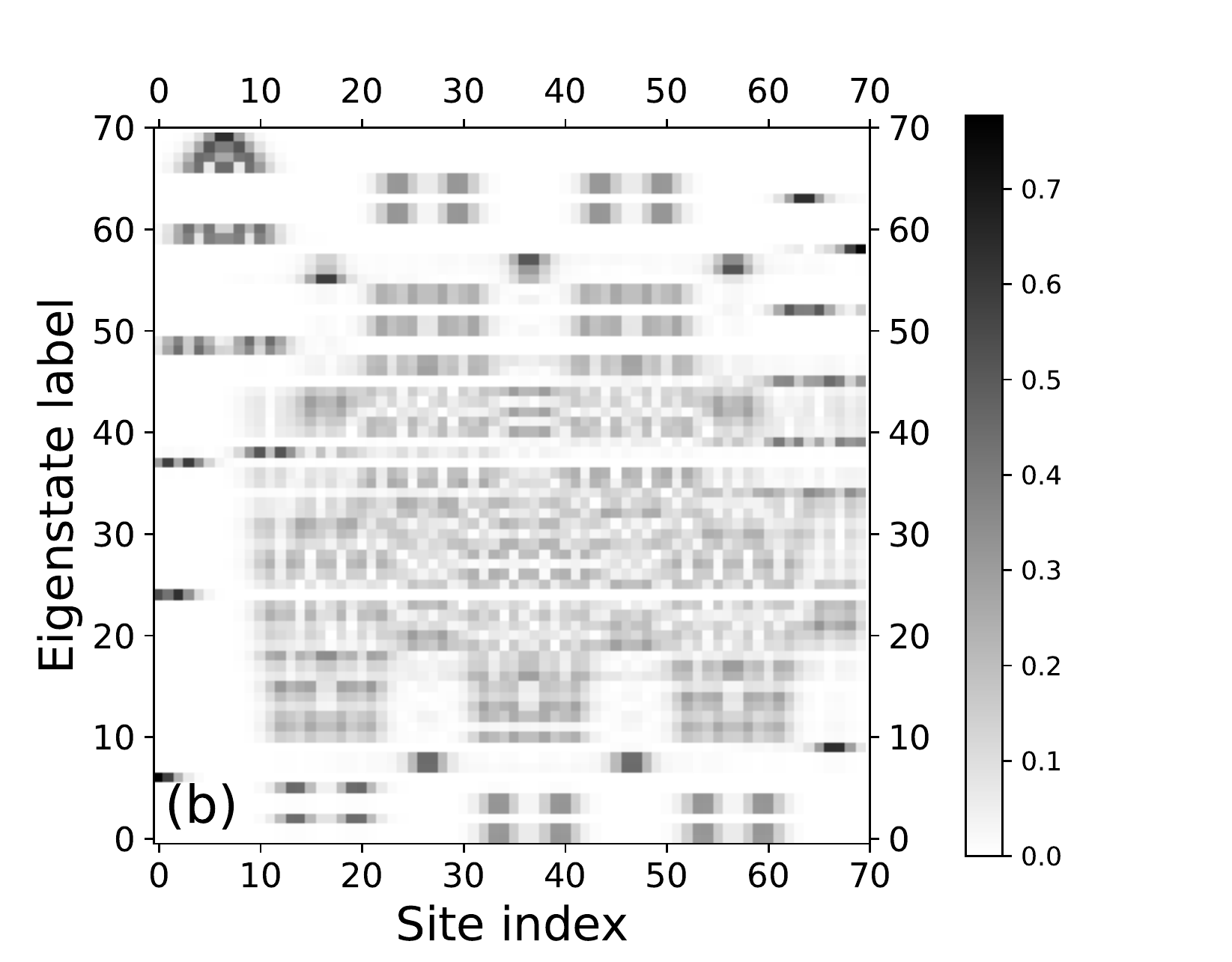}
\caption{(a) Spectrum of the energy eigenvalues for a TB chain emerging from a seed ABCDEFG and subsequent application
of the 7:3 LSD up to a total of 70 sites, which includes three unit-cells of the final periodic behaviour.
The line is drawn to guide the eye.
(b) The corresponding grey scale eigenstate map showing the magnitude of the amplitudes on all
sites of the chain for all eigenstates whose labeling $0,...,69$ is according to an increasing eigenenergy.
We use open BCs and the on-site energies corresponding to $\{$A,B,C,D,E,F,G$\}$ 
are $\{1,2,3,4,5,6,7\}$ respectively whereas the off-diagonal coupling is equal to one.}
\label{fig:Fig4ab}
\end{figure}

The LSD chain following the 7:3 rule possesses a transient of ten chain elements
before it becomes periodic with period 20 and with the unit-cell DCBAABCCBAABCDDCBBCD (see table \ref{tab:table1}).
This unit-cell contains now only the symbols A,...,D i.e. the high on-site energies according to E,F,G have been removed in the
course of the LSD operations. Fig.\ref{fig:Fig4ab}(a) shows the spectrum of the energy eigenvalues for the 7:3 LSD chain.
Compared to the 7:1 chain of the previous subsection the plateau-like clustering of the energy eigenvalues is
attenuated in the central part of the energy eigenvalue spectrum at the cost of an almost linear envelope behaviour.
For the regimes of low and high energies this plateau behaviour is still pronounced. Remarkably, at very high energies
a significant enhancement of the spacing among the eigenvalues and consequently of the slope of the spectral 
eigenvalue curve is observed.

Fig. \ref{fig:Fig4ab}(b) shows the eigenstate map for the 7:3 case which defers quite significantly from the corresponding
one of the 7:1 case, also from the 7:2 case not shown here.
For the lowest energies the eigenstates 0-5 are predominantly localized on the local reflection symmetric BAAB subparts of the chain
which are centered around the site pairs $13/14$ and $19/20$ and repeated after 20 further sites according
to the length of the unit-cell. The localization centered on the first two pairs $13/14$ and $19/20$ in the chain possess a slightly
higher energy as compared to the following ones due to their tail extending onto the high on-site energies of some
of the seed sites. State 6 is a left edge localized state similar to state 24 and 37. States 7 and 8 are localized
on the local reflection symmetric CBBC part of the chain involving sites with a higher on-site energy and state 9 is centered on the same
sequence but possesses a higher energy due to its position at the right edge. With further increasing degree
of excitation the BAAB centered states further delocalize and spread. From state 19 on an energetically 
broad band of delocalized states appears which resides in particular on the subsequences CBAABCCBAABC possessing also a 
high degree of local symmetry. States 19 to 23 and 25 to 36 as well as 39 to 45 are completely delocalized
on the sequence of the three involved unit-cells. For even higher energies eigenstate localization sets in again
with a mix of centering around the BCCB, CBBC and CDDC parts of the chain. On top of this behaviour the occupation
of the sites of the original seed and its first reflection A,....G,G,...A in terms of localized eigenstates appears
for the high energy states 48,49 as well as 59,60 and finally a series of states 66-69 increasingly narrow down their amplitudes
around the center sites GG. 

We now analyze the 7:5 LSD chain which possesses a transient of 32 sites
before it becomes periodic with period 24 with the unit-cell BAABBAABBAAAABBAABBAABBB (see table \ref{tab:table1})
consisting of two different possibilities for the on-site energies.  
Fig.\ref{fig:Fig4ab}(a) shows the spectrum of the energy eigenvalues for the 7:5 LSD chain.
The trend realized in the above discussion of the 7:3 chain continues further i.e. we have an energetically
central part of the spectrum for which an approximately linear envelope behaviour can be observed.
At low energies the plateau-like clustering still persists. For higher energies we observe
an extended plateau before a steep rise of the energies occurs due to an, in part, enhanced energy spacing.

\begin{figure}[t]
\hspace*{-1.1cm} \includegraphics[width=0.8 \columnwidth]{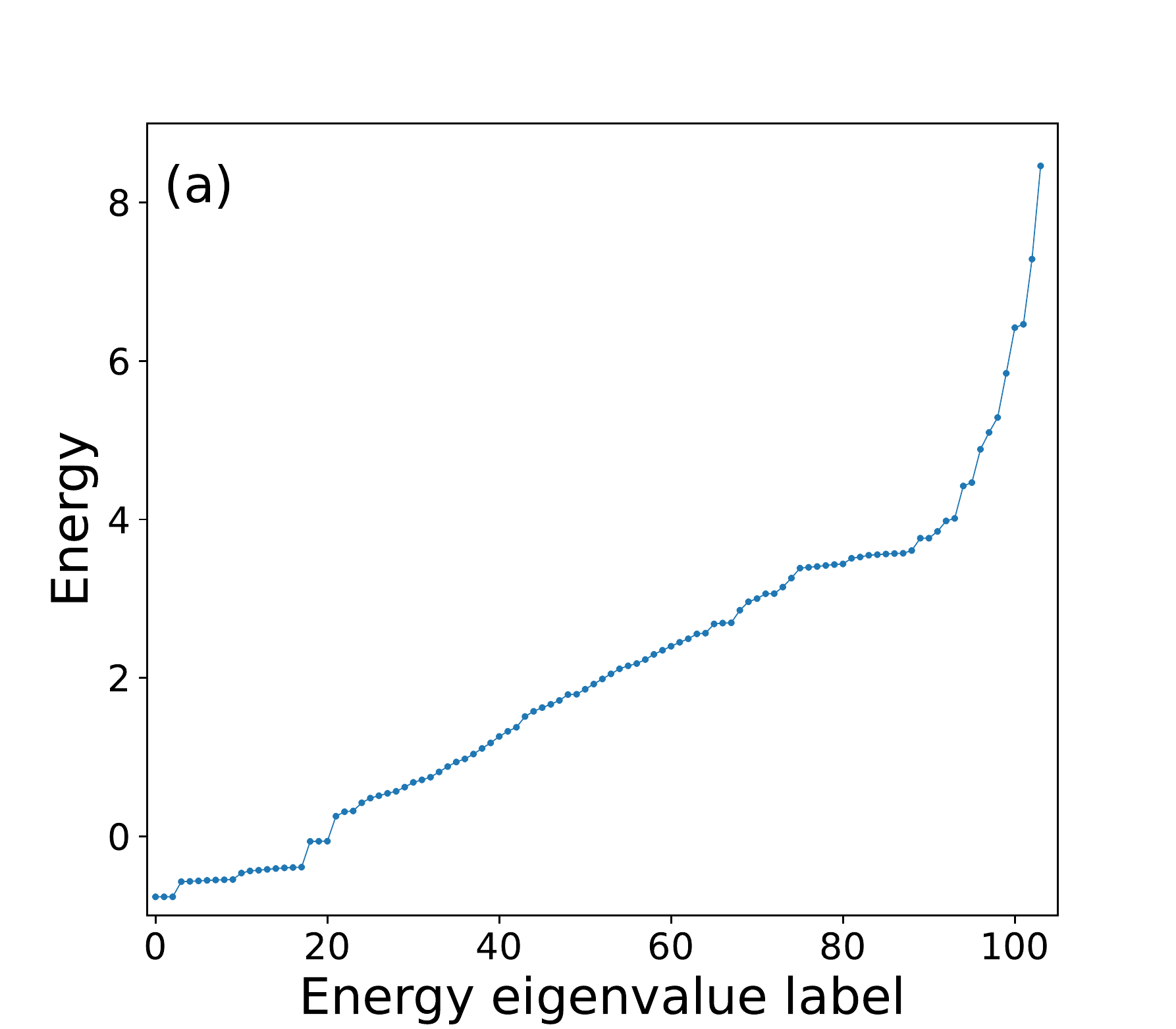} 
\includegraphics[width=\columnwidth]{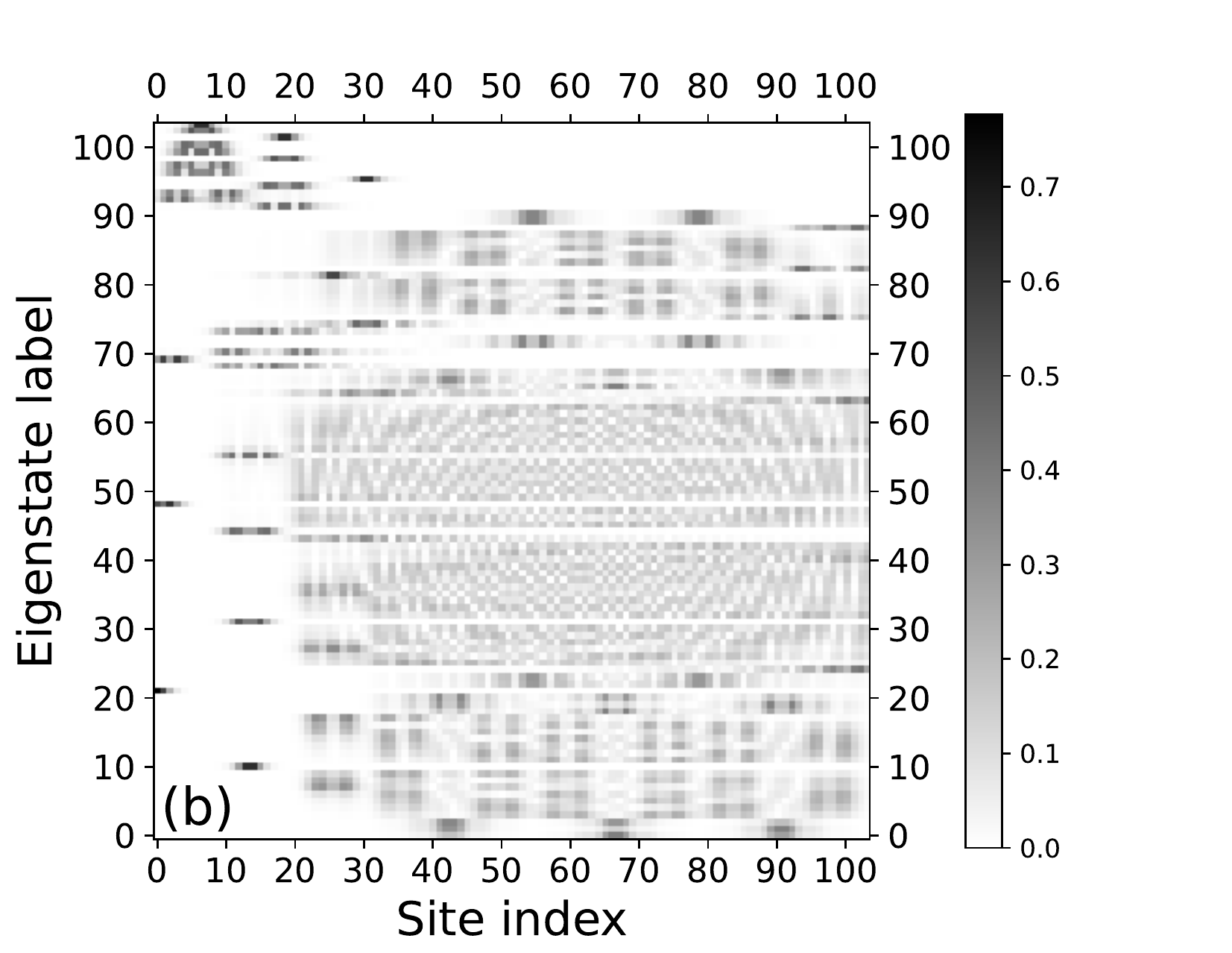} 
\caption{(a) Spectrum of the energy eigenvalues for a TB chain emerging from a seed ABCDEFG and subsequent application
of the 7:5 LSD up to a total of 104 sites, which includes three unit-cells of the final periodic behaviour.
The line is drawn to guide the eye.
(b) The corresponding grey scale eigenstate map showing the magnitude of the amplitudes on all
sites of the chain for all eigenstates whose labeling $0,...,103$ is according to an increasing eigenenergy.
We use open BCs and the on-site energies corresponding to $\{$A,B,C,D,E,F,G$\}$ 
are $\{1,2,3,4,5,6,7\}$ respectively whereas the off-diagonal coupling is equal to one.}
\label{fig:Fig5ab}
\end{figure}

Fig. \ref{fig:Fig5ab}(b) shows the eigenstate map for the 7:5 case. Due to the presence of only two on-site energies
and the occurence of multiple locally symmetric sequences such as BAAB, BAAAAB, as well as ABBBBA subsequences
there is a larger portion of delocalized eigenstates observable that however do not populate the original seed
sites. Remarkable are also the progression of localized
states centered around the sites 13/14 and for high energies centered around the GG and the DEED sequence which
increasingly narrow down to their central sites with increasing energy.

\subsection{The 7:6 LSD chain}

\begin{figure}[t]
\hspace*{-1.1cm} \includegraphics[width=0.8 \columnwidth]{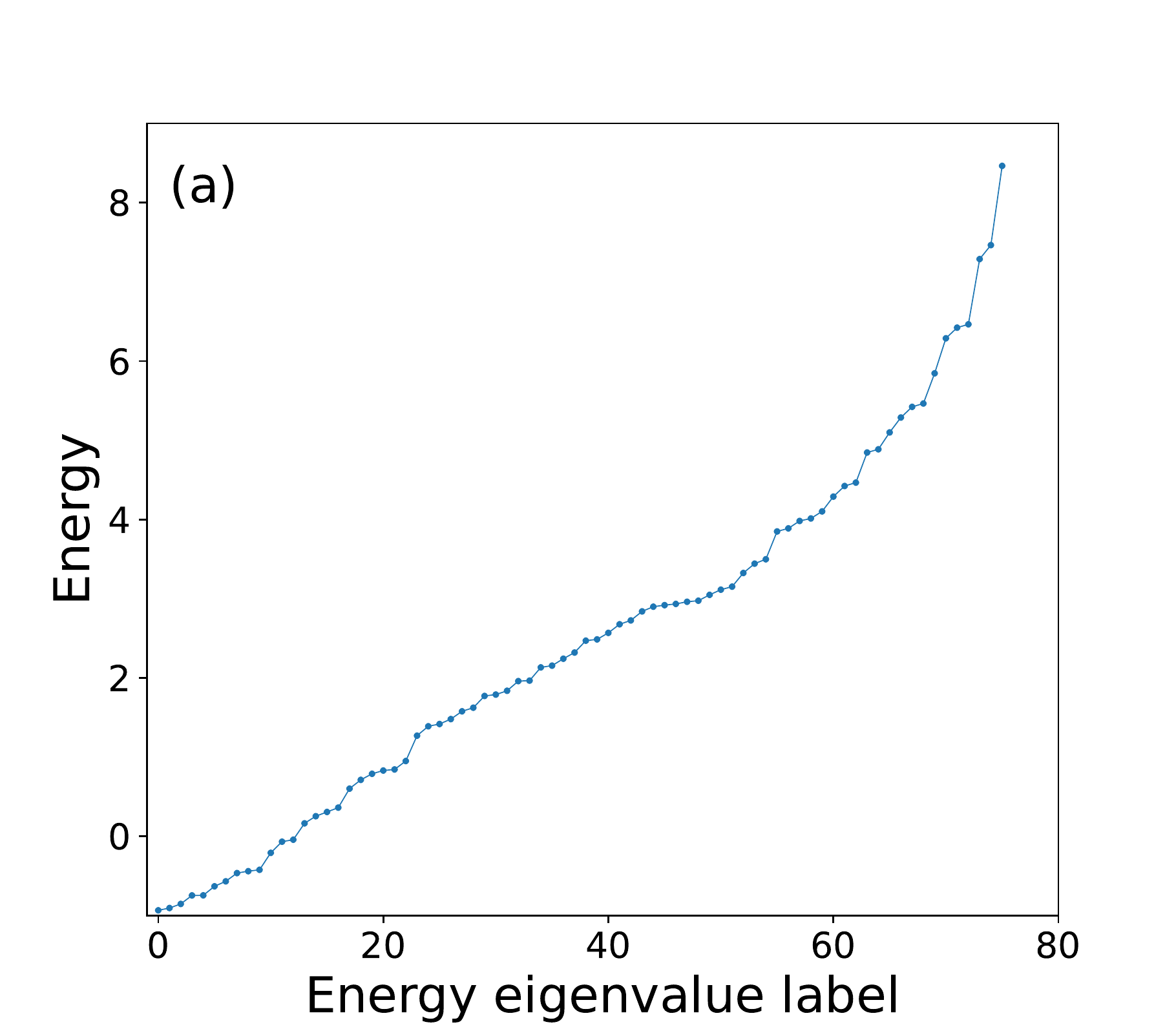} 
\includegraphics[width=\columnwidth]{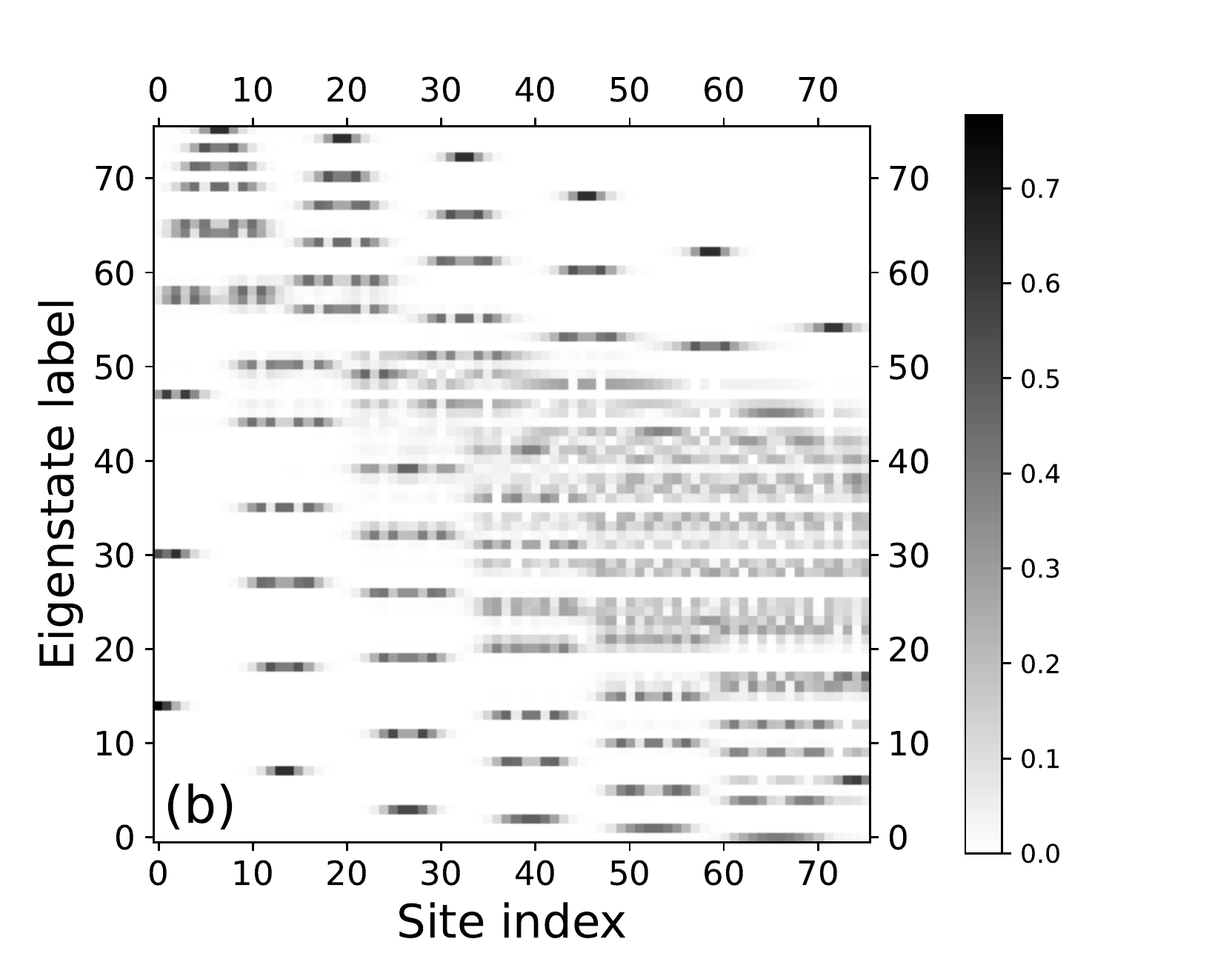} 
\caption{(a) Spectrum of the energy eigenvalues for a TB chain emerging from a seed ABCDEFG and subsequent application
of the 7:6 LSD up to a total of 76 sites, which includes three unit-cells of the final periodic behaviour.
The line is drawn to guide the eye.
(b) The corresponding grey scale eigenstate map showing the magnitude of the amplitudes on all
sites of the chain for all eigenstates whose labeling $0,...,75$ is according to an increasing eigenenergy.
We use open BCs and the on-site energies corresponding to $\{$A,B,C,D,E,F,G$\}$ 
are $\{1,2,3,4,5,6,7\}$ respectively whereas the off-diagonal coupling is equal to one.}
\label{fig:Fig6ab}
\end{figure}

The 7:6 LSD chain possesses a transient of 73 sites
before it becomes periodic with period 1 with the unit-cell A (see table \ref{tab:table1})
consisting of only a single possibility for the on-site energies.
Fig.\ref{fig:Fig6ab}(a) shows the spectrum of the energy eigenvalues for the 7:6 LSD chain.
Compared to the previous cases an approximately linear envelope behaviour is now present also for the low energies and for
high energies a smooth nonlinear upward is observed.

The resulting eigenstate map in Fig.\ref{fig:Fig6ab}(b) shows several series of localized eigenstates in a progressive manner in the 
low to intermediate energy regime. This is a remarkable localization behaviour and happens at the cost of few delocalized states
as compared to the previously discussed cases 7:x, x=1,3,5. The origin of these unexpected localization features
are again the locally reflection symmetric sequences occuring in the chain, but now they come along with a certain transient
scaling behaviour. This means that along the chain (see table \ref{tab:table1}) we encounter now the subsequences
$\mathrm{B(A)_2B}$, $\mathrm{B(A)_4B}$, $\mathrm{B(A)_6B}$, $\mathrm{B(A)_8B}$ and $\mathrm{B(A)_{10}B}$ centered around the
sites 13/14, 26/27, 39/40, 52/53 and 65/66 respectively. Here $(\mathrm{A})_n$ stands for an $n-$fold repeated symbol A.
Naturally, the width of these localized eigenstates increases with increasing number of sites involved in the above-mentioned
sequences $\mathrm{B(A)_nB}$. In the high energy regime another several series of localized eigenstates appear which are
now centered and localized on the sequences FGGF, EFFE, DEED, CDDC and BCCB with decreasing energy. Their widths are comparable
as can be observed in Fig.\ref{fig:Fig6ab}(b). 

\begin{figure}[t]
\hspace*{-0.3cm} \includegraphics[width=\columnwidth]{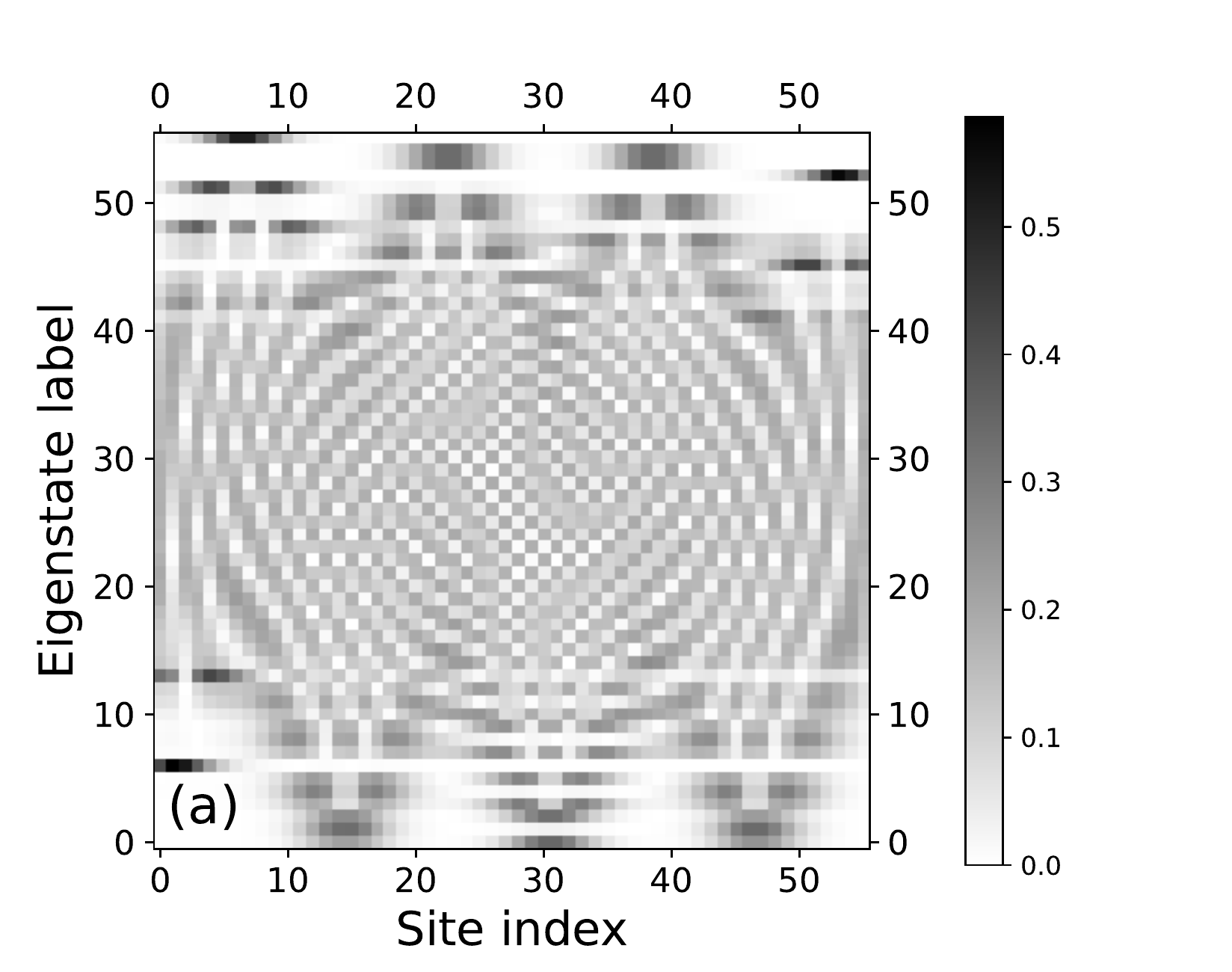} 
\includegraphics[width=\columnwidth]{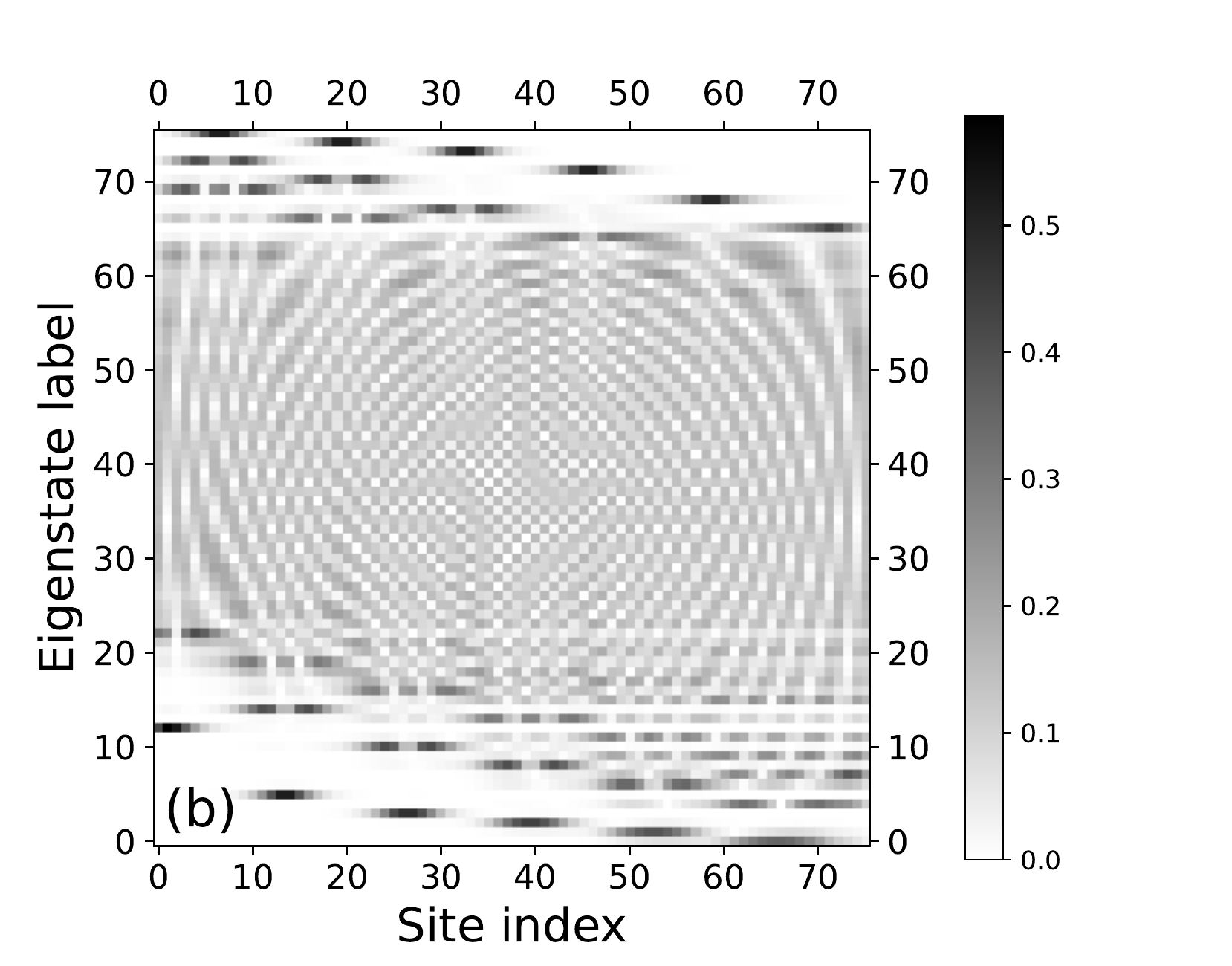} 
\caption{(a) Grey scale eigenstate map for a TB chain with strong coupling $C=5$ emerging from a seed ABCDEFG and subsequent application
of the (a) 7:1 LSD up to a total of 56 sites, which includes three unit-cells of the final periodic behaviour.
(b) The same like in (a) but for the 7:6 LSD chain up to 76 sites. We use open BCs and the on-site energies
corresponding to $\{$A,B,C,D,E,F,G$\}$ are $\{1,2,3,4,5,6,7\}$ respectively.}
\label{fig:Fig7ab}
\end{figure}

\subsection{LSD chains with stronger off-diagonal couplings} \label{subsec:sc}

Let us finally explore the case of an enhanced off-diagonal coupling strength $C=5$. 
Fig.\ref{fig:Fig7ab} shows the eigenstate map for the TB chain for such a strong coupling for the cases of an (a) 7:1 and (b) 7:6
LSD rule. Compared to the case of a weaker coupling discussed in the previous subsections we observe now an energetically
broad band for intermediate energies for which the eigenstates are delocalized over the complete chain, including the
initial seed of the chain. For some comparatively narrow regions of high and low energies series of localized eigenstates
appear whose amplitudes are centered around the locally symmetric subparts of the chains, as discussed above. The broadening (for
low energies) respectively narrowing (for high energies) of the localization of the eigenstates with increasing degree of excitation
can also be observed here. In Fig.\ref{fig:Fig7ab}(b) there is a series of domain localized eigenstates 0-4 that possess
a small but finite overlap and therefore covers, apart from the initial seed, the complete chain. Similarly there is a series of
slightly overlapping eigenstates in the high energy regime that cover the complete chain.

\section{Conclusions and Outlook} \label{sec:concl}

The present work addresses a class of systems, specifically one-dimensional chains, which fall into the
gap between highly ordered periodic crystals as well as quasicrystals with aperiodic long-range order
and disordered chains. On an abstract level, they are constituted from a symbolic seed sequence by the repeated 
application of local reflection operations following certain rules. We have been here focusing on the
class of rules $n:m$ which state that the local reflection operations involve $n$ and $m$ symbols in an alternating
periodic manner. A study of a few example cases showed us that the repeated application of the local reflection operation
provides us, starting from the seed, via a transient behaviour with an 'asymptotic' periodic behaviour. The
content i.e. appearance of the resulting unit-cell largely varies with the rules $n:m$. We have provided
a general proof of this behaviour thereby understanding the variability of the pathway to periodicity
and its final appearance. As a consequence, and due to the constructive character of the proof,
we could predict the final period, the final number of independent chain elements and the overall appearance of the
unit-cell. This appearance is composed of an inhomogeneous alternating series of subchains with
local symmetries, and of varying length and recursively defined content.

As a consequence of the construction principle of the chain it contains
a large number of local reflection and translation symmetries that strongly overlap and repeat in
different shapes and content.
Opposite to e.g. the substitution rules generating aperiodic long-range ordered
chains our local symmetry dynamics rule leaves intact the chain of a certain generation while building
up the next one. 
The strong impact of the presence of these multiple and nested local
symmetries on the properties of such a chain became very clear in the second part of this investigation.
We mapped the symbolic code onto a tight-binding Hamiltonian where the on-site energies follow
the chains symbolic sequence in the form of corresponding numerical values. In the course of a corresponding
spectral analysis, including the behaviour of the energy eigenvalues and the eigenstates, a rich
variability could be observed. Specifically the localization properties of the eigenstates 
changed from series of localized eigenstates whose localization centers are determined by
the locally reflection symmetric subchains to delocalized states sometimes in a continuous and
sometimes in a more abrupt manner with varying excitation energy. Our analysis clearly demonstrates that local
and in particular overlapping local symmetries provide a powerful means to control the centering and spreading
of eigenstates.
Certain locally symmetric domains act as a nucleus for the increased spreading of localized eigenstates
with increasing degree of excitation whereas others show a corresponding inverted behaviour.
Depending on the assigned on-site values to the domains and their
couplings they appear in the low or high energy regime with a band of delocalized eigenstates separating them.
This opens the pathway of a systematic design of the localization of the eigenstates individually
and with respect to each other by the sophisticated use of the overlapping local symmetry domains.

While we have been focusing in the present work on a specific class of local symmetry operations that generate the chain
there is many other possibilities and open question to be addressed in the future. Reaching beyond the
$n:m$ rule can be imagined in many different ways. One problem is the roadmap to periodicity: under what conditions
does periodicity occur and if not, what are the possible asymptotic behaviours passing the transient phase ?
Is there new classes of order that possibly emerge via local symmetry dynamics generated chains and
what would be their spectral or even topological properties ? These, and related questions, are left
to future investigations.

\vspace*{1cm}

\section{Acknowledgments} \label{sec:acknowledgments}

The author acknowledges discussions with F.K. Diakonos, M. R\"ontgen and M. Pyzh.
This work is supported by the Cluster of Excellence “Advanced Imaging of Matter” of the Deutsche
Forschungsgemeinschaft (DFG)-EXC 2056, Project ID No. 390715994.

\end{document}